\documentclass[aps,prl,twocolumn,amsmath,amssymb, superscriptaddress,floatfix]{revtex4-2}

\usepackage{amsmath}    
\usepackage{amssymb}
\usepackage{graphicx}   
\usepackage{verbatim}   
\usepackage{color}      
\usepackage{hyperref}   
\usepackage{blindtext}
\usepackage{natbib}
\usepackage{fixmath}
\usepackage{enumitem}
\usepackage{dsfont}

\newcommand{\ket}[1]{\left| #1 \right\rangle}

\begin{document}

\title{Loss-induced quantum information jet in an infinite temperature Hubbard chain}

\author{Patrik Penc}
\affiliation{Department of Theoretical Physics, Institute of Physics, Budapest University of Technology and Economics, M\H{u}egyetem rkp. 3., H-1111 Budapest, Hungary}
\affiliation{HUN-REN–-BME Quantum Dynamics and Correlations Research Group, Budapest University of Technology and Economics, M\H{u}egyetem rkp. 3., H-1111 Budapest, Hungary}
\affiliation{Strongly Correlated Systems 'Lend\" ulet' Research Group, HUN-REN Wigner Research Centre for Physics, P.O. Box 49, 1525 Budapest, Hungary}
\author{C\u at\u alin Pa\c scu Moca}
\affiliation{HUN-REN–-BME Quantum Dynamics and Correlations Research Group, Budapest University of Technology and Economics, M\H{u}egyetem rkp. 3., H-1111 Budapest, Hungary}
\affiliation{Department of Physics, University of Oradea, 410087, Oradea, Romania}

\author{\"Ors Legeza}
\affiliation{Strongly Correlated Systems 'Lend\" ulet' Research Group, HUN-REN Wigner Research Centre for Physics, P.O. Box 49, 1525 Budapest, Hungary}
\affiliation{
Institute for Advanced Study,Technical University of Munich, Germany, Lichtenbergstrasse 2a, 85748 Garching, Germany
}

\author{Toma\v z Prosen}
\affiliation{Department of Physics, Faculty of Mathematics and Physics,
University of Ljubljana, Jadranska 19, SI-1000 Ljubljana, Slovenia}

\affiliation{Institute for Mathematics, Physics, and Mechanics, Jadranska 19, SI-1000 Ljubljana, Slovenia}

\author{Gergely Zar\'and}
\affiliation{Department of Theoretical Physics, Institute of Physics, Budapest University of Technology and Economics, M\H{u}egyetem rkp. 3., H-1111 Budapest, Hungary}
\affiliation{HUN-REN–-BME Quantum Dynamics and Correlations Research Group, Budapest University of Technology and Economics, M\H{u}egyetem rkp. 3., H-1111 Budapest, Hungary}

\author{Mikl\'os Antal Werner}
\email{werner.miklos@wigner.hun-ren.hu}
\affiliation{Department of Theoretical Physics, Institute of Physics, Budapest University of Technology and Economics, M\H{u}egyetem rkp. 3., H-1111 Budapest, Hungary}
\affiliation{HUN-REN–-BME Quantum Dynamics and Correlations Research Group, Budapest University of Technology and Economics, M\H{u}egyetem rkp. 3., H-1111 Budapest, Hungary}
\affiliation{Strongly Correlated Systems 'Lend\" ulet' Research Group, HUN-REN Wigner Research Centre for Physics, P.O. Box 49, 1525 Budapest, Hungary}

\date{\today}
\begin{abstract}
 Information propagation in the one-dimensional  infinite temperature Hubbard model with a dissipative particle sink at the end of a semi-infinite chain is studied. In the strongly interacting limit, the two-site mutual information and the operator entanglement entropy exhibit a rich structure with  two propagating information fronts and superimposed interference fringes.  
 A classical reversible cellular automaton model quantitatively captures  the transport  and  the slow,  classical part of the correlations, but fails to describe the rapidly propagating information jet.  The fast  quantum jet resembles coherent free particle propagation, with the accompanying long-ranged interference fringes that are exponentially damped by short-ranged spin correlations in the many-body background.
\end{abstract}

\maketitle

\emph{Introduction.\,– }The creation and the fate of classical and quantum correlations and  information in open quantum systems is  of tremendous  importance  in  quantum  technologies and quantum information science~\cite{Zurek-2003,Horodecki-2005, Breuer-2016}. Coupling a quantum system to the environment is a necessity for initializing quantum states, but it is also inevitable during  manipulation and time evolution, when it leads to decoherence.
By now various quantum platforms exist including cold atomic systems~\cite{Diehl-2008, Muller-2012, Bloch-2017}, superconducting and photonic quantum-computational platforms~\cite{Preskill-2018, Arrazola-2021, GoogleQAI23}, and quantum tweezer arrays~\cite{Anderegg-2019, Ma-2022}, which allow us to design open quantum systems and study their behavior with unprecedented control and insight.

Quantum states of open systems are generically mixed, and host correlations of both quantum and classical nature~\cite{Modi-2010,Berrada-2012, Adesso-2016}. Distinguishing these two, if possible at all, can usually be done only for very small systems~\cite{Maziero-2009,Berrada-2012}, or for non-interacting models~\cite{Adesso-2010}.
Disentangling them and shedding light on the non-equilibrium information transport and correlations in large, interacting many-body quantum systems is therefore of great interest~\cite{Harbola-2008,Prosen-2010, Daley-2014, Aolita-2015, Malouf-2020, Weimer-2021, Bertini-2021}. 

\begin{figure}[b]
    \includegraphics[width=1\columnwidth]{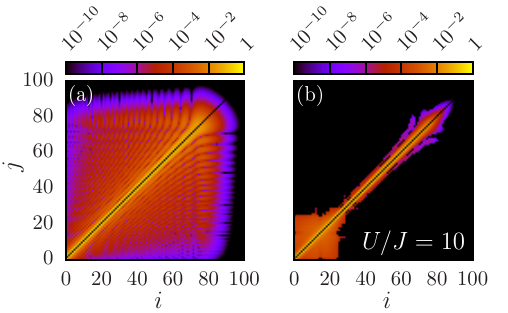}
    \caption{Two-site mutual information \eqref{eq:MI} at time $t J=80$ in (\textit{a}) the non-interacting limit, and (\textit{b})  for  $U/J=10$  and a dissipator strength $\Gamma/J = 0.5$. For  strong interactions, $U/J=10$, long-range correlations  are visible  at sites closer to the sink ($i,j \lesssim 30$), while a jet signal reaches $i,j \approx 80$.}
    \label{fig:MI_fig1}
\end{figure}

In this letter, we study information and operator space entanglement entropy propagation~\cite{Prosen-2007, Alba-2019} in  the one-dimensional infinite temperature Hubbard model~\cite{Essler-2005}
\begin{equation}
    {H} = -\frac{J}{2} \sum_{i=1}^{L-1} \sum_{\sigma = \uparrow, \downarrow} \left(c^\dag_{i\,\sigma} c_{i+1\,\sigma}  + h.c.\right) + U \sum_{i}^L n_{i \uparrow} n_{i \downarrow} \; , \label{eq:Hubbard}
\end{equation}
with a dissipative particle sink  attached to the left end of the chain. Here  $c_{i \sigma}^\dag$($c_{i \sigma}$) denote  fermionic creation (annihilation) operators, $J$ is the hopping, and  $U$ the interaction. A dissipative particle sink (described later) is introduced at the first site within a Lindbladian approach~\cite{Lindblad-1976,Gorini-1976}. 
Note also that a related similar setup of localized particle source has very rich behavior even for free fermions/bosons~\cite{Spohn,Sels}, and at low temperatures~\cite{Froml-2019,Froml-2020,Wolff-2020}.

\begin{figure*}
    \centering
    \includegraphics[width=\textwidth]{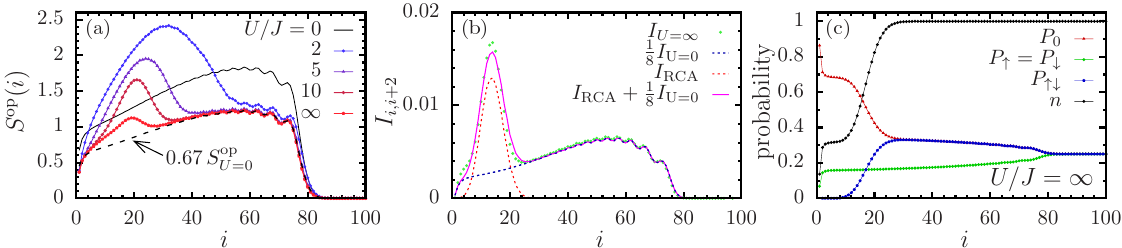}
    \caption{(\textit{a}) Operator  entanglement entropy $S^\mathrm{op}(i)$ at time $t J=80$ and dissipator strength $\Gamma/J = 0.5$ for various interaction strengths. A clear double-front structure builds up for increasing interaction strength. The  faster front travels at a speed unaltered by  interactions, and it is accompanied by interference fringes.  The velocity of the second front is suppressed 
    with increasing $U$, and converges to a finite value as  $U\to\infty$. At strong interactions, $U/J \gtrsim 5$, the fast entropy 
    profile is identical to    $0.67$-times  the non-interacting value. (\textit{b}) Two-site mutual information $I_{i, \,i+2}$ 
     for $U=\infty$, $\Gamma/J=0.5$.  $I_{i, \,i+2}$ can be decomposed into an incoherent classical mutual information, $I_{RCA}$, perfectly captured by a reversible cellular automaton (RCA)  model,  and to a non-interacting profile  reduced by a factor ${1}/{8}$.
     (\textit{c}) Position dependent  probabilities of the states $\ket{0}$, $\ket{\uparrow}$, $\ket{\downarrow}$, and $\ket{\uparrow \downarrow}$ . The uniform distribution is broken at the fast front, but particle-hole symmetry is maintained,  $P_0 = P_{\uparrow \downarrow}$. }
    \label{fig:op_entropy}
\end{figure*}

 The system is initialized in a half-filled, infinite temperature state\, and the sink is turned on at time $t=0$ instantaneously~\cite{Moca-2022}. A rich picture of correlations  appears at times $t>0$ (see Fig.~\ref{fig:MI_fig1} for typical results). In the absence of interactions, holes propagate coherently from the sink with a velocity $v_0 = J$, and the mutual information displays characteristic  interference fringes. In contrast, for strong interactions, two clearly distinguishable ballistically propagating fronts appear. Long-ranged correlations build up only after the slower front. 
 As we demonstrate, these  can be well-understood through a simple classical charged hard-point gas reversible cellular automaton (RCA)~\cite{RCA1,RCA2,RCA3}. Simultaneously, a fast, short-range correlated jet signal also builds up and propagates with the unrenormalized speed $v_0=J$, signaling the presence of  fast, coherent quantum excitations. The  operator entanglement profile exhibits  a similar two-front structure (see Fig.~\ref{fig:op_entropy}.a), bearing resemblance to the  transport profiles observed in one-dimensional integrable models~\cite{Scopa-2021,Scopa-2022}.

\emph{Numerical methods. –}
We determine the time-de\-pen\-dent density operator, ${\rho}(t)$, by the numerical solution of the Lindblad equation~\cite{Gorini-1976, Lindblad-1976},
\begin{equation}
    i \frac{d  \rho}{d t} 
    = [{H},{\rho}] +  i\, \Gamma \sum_\sigma \bigl(
    2 \,{c}_{1\sigma} {\rho} \,{c}_{1\sigma} ^\dag- \bigl\lbrace {c}_{1\sigma}^\dag  {c}_{1\sigma} , {\rho} \bigr\rbrace \bigr) \; . \label{eq:Lindblad}
\end{equation}
Here the first term generates the usual von Neumann evolution, governed by  ${H}$, while the second 'dissipator' term accounts for the particle sink with rate $\Gamma$. Studying the effect of localized and controlled particle loss is also motivated by experiments with  cold atoms~\cite{Barontini-2013,Labouvie-2016}, where a well-focused electron beam creates the dissipative sink. The specific position of the sink is chosen to keep the required chain length hence the numerical resources low, while this position is not expected to qualitatively influence the observed physics.

Numerical simulation of \eqref{eq:Lindblad} is a challenging problem due to the exponential dimensionality of the Liouville space of density operators. Here we use the matrix product state method~\cite{White-1992, Vidal-2004, Schollwock-2011} applied to open systems~\cite{Verstraete-2004, Zwolak-2004, Prosen-2009}. To further reduce the dimensionality of the problem, however, we make use of the non-Abelian approach  of Refs.~\cite{Werner-2020,Moca-2022}, and exploit the global $\mathrm{SU}(2) \times \mathrm{U}(1)$ symmetry of the Liouvillian, also preserved by the dissipators. (See also Ref.~\cite{Albert-2014})
 Simulations can also be extended to the $U=\infty$ limit by suppressing all processes that change the total interaction energy. In this limit, not only the total number of fermions but also the total number of doubly occupied sites is conserved, which is enforced in the simulations as a constraint.

\emph{Mutual information. –}
Non-equilibrium evolution of charge densities and currents  have been thoroughly investigated in our setup in Ref.~\cite{Moca-2022}. Here we focus on the evolution of  information and entanglement measures. 

One- and two-site observables can be determined by the one- and two-site reduced density matrices, ${\rho}_i = \mathrm{Tr}_{L \setminus \lbrace i \rbrace} {\rho} $ and ${\rho}_{ij} =  \mathrm{Tr}_{L \setminus \lbrace i,j \rbrace} {\rho}$, with $\mathrm{Tr}_{L \setminus X}$ the partial trace over of all  sites excepting the ones  in the set $X$~\cite{Legeza-2006}. 
The total strength of correlations between two lattice sites is quantified by the  mutual information~\cite{Groisman-2005, Luo-2012, Barcza-2015},
\begin{equation}
    I_{ij} = S_i + S_j - S_{ij} \; , \label{eq:MI}
\end{equation}
where $S_X = - \mathrm{tr} {\rho}_X \log {\rho}_X$ stands for the von Neumann entropy of
subsystem  $X$. 

In the non-interacting case, we can use the third quantization method of Ref.~\cite{Prosen-2008} to compute $I_{ij} $ analytically (panel (a) in Fig.~\ref{fig:MI_fig1}). In this case, $I_{ij}$ propagates with a speed $v_0= J$, the maximal speed of free fermions, and characteristic correlation fringes are produced for $i,j\lesssim t J$.  The edge of these fringes follows the depletion front, created by the sink. 

The situation changes, however, dramatically for $U/J\gg 1$ (panel (b) in Fig.~\ref{fig:MI_fig1}): 
then the depletion region propagates with a suppressed speed~\cite{Moca-2022}. 
A mutual information pattern develops in this slowly moving depletion region, 
while interference fringes are, however,  suppressed. As we argue later, 
correlations here are essentially classical.  However, outside the density front, 
a fast and narrow mutual information jet emerges, which propagates with 
unrenormalized velocity $v_0$, and is characterized by 
coherent oscillations and interference patterns.
 Fig.~\ref{fig:op_entropy}.b displays  the spatial dependence of the mutual information $I_{i,i+2}$, which demonstrates the presence of two distinct parts: a slow, classical contribution, explained later in terms of classical RCA, and a coherent quantum contribution, associated with the motion of spinless fermionic excitations.

\emph{Operator entanglement. –}
Similar double-front features appear in the so-called operator  entanglement entropy~\cite{Prosen-2007} (see Fig.~\ref{fig:op_entropy}),
defined through the Schmidt decomposition of the density matrix,
$ {\rho} = \sum_{\alpha} \lambda^{(i)}_\alpha {A}^{(i)}_\alpha \otimes {B}^{(i)}_\alpha$, where  the (normalized) Schmidt-operators 
${A}^{(i)}_\alpha$ and ${B}^{(i)}_\alpha$ act on lattice sites $\lbrace 1,2,\dots i \rbrace$ and $\lbrace i+1, \dots L \rbrace$, respectively. 
We  define the operator entanglement entropy  from the Schmidt-coefficients 
by introducing  the probabilities $  p^{(i)}_\alpha \equiv |\lambda^{(i)}_\alpha|^2/ (\sum_\alpha |\lambda^{(i)}_\alpha|^2)$,  
\begin{equation}
    S^{\mathrm{op}}(i) = - \sum_\alpha p^{(i)}_\alpha \log p^{(i)}_\alpha \; .  
\end{equation}

Our results on the operator entanglement entropy are summarized in Fig.~\ref{fig:op_entropy}.a
 for various interaction strengths at fixed, long time, $tJ=80$. 
All presented curves, excepting $U=0$, display two fronts. Strikingly, the velocity and overall shape of the faster front is almost not affected by the interaction, while its amplitude is reduced by a factor $\gamma \approx 0.67$. This provides indirect proof that the fast front is carried by a coherent fermionic quasiparticle, which moves through the infinite temperature state coherently, even in the presence of strong interactions.
While an almost precise agreement between the non-interacting and interacting profiles is observed for intermediate dissipator strengths, with a prefactor being close to $2/3$, the value of the prefactor and the shape of the curves get slightly distorted for small and large $\Gamma$ values~\cite{suppmat}, where the outflowing current is also suppressed~\cite{Moca-2022}. In contrast to the fast front, the second front is slowed down by interactions but, remarkably, its velocity remains finite even in the $U=\infty$ limit. Furthermore, the velocity of this slower front is found to depend also on the total current flowing to the sink~\cite{suppmat}. 

\emph{Occupation profiles. –}
Fig.\ref{fig:op_entropy}.c shows the spatial dependence of the probabilities of the states $|0\rangle$, $|\uparrow\rangle$, $|\downarrow\rangle$, and $|\uparrow \downarrow\rangle$ 
at time $t=80$, as obtained by $U=\infty$ quantum simulations. A clear double-front structure appears also in the occupations, with the equality $P_\uparrow(i) = P_\downarrow(i)$ satisfied due to the residual SU(2) spin symmetry of the Lindbladian.
 A fast, particle-hole symmetric signal with $P_{\uparrow \downarrow} = P_0 > \frac{1}{4}$ with exponential accuracy, and $P_\uparrow = P_\downarrow < \frac{1}{4}$ propagates with the maximal velocity of free fermions, $v$, leaving the total occupation number unchanged, $n(i) = 1$.
 As this fast front  remains particle-hole symmetric, it remains invisible in 
 charge and current densities. 
  In contrast, before the second, slower front, the probabilities $P_{\uparrow\downarrow}$ and $P_0$ become different, and a charge density profile develops.

\emph{Reversible Cellular Automaton Model. –}
RCA models attracted significant interest in the context of integrable systems in recent years~\cite{Prosen-2016,Alba-2019,Gombor-2022}, since they can capture many aspects 
of classical Hamiltonian dynamics and, in certain cases,  their simplicity allows for exact steady-state or time-dependent solutions~\cite{Alba-2019,RCA1,RCA2,RCA3,Klobas-2021}.
 Here we show that a significant part of the features observed in our quantum simulations  in the large $U$ limit are captured by type $(2+2)$ XXC cellular automaton model~\cite{Gombor-2022}.
  In this model, each site has four possible states,
  $\lbrace 0,\, \uparrow,\, \downarrow,\, \uparrow\downarrow \rbrace$, and the two-site local update rules are 
  transmissions $(0, \sigma) \Leftrightarrow (\sigma,0)$, $(\uparrow \downarrow, \sigma) \Leftrightarrow ( \sigma,\uparrow \downarrow)$ and reflections  $(\sigma, \bar\sigma) \Leftrightarrow (\sigma,\bar \sigma)$, $(0, \uparrow \downarrow) \Leftrightarrow (0, \uparrow \downarrow)$, where $\sigma \in \lbrace \uparrow, \downarrow \rbrace$ is used for the spin of a singly occupied site. 
 Exchange processes such as $(\uparrow \downarrow, 0) \Leftrightarrow ( 0,\uparrow \downarrow)$ or 
  $(\sigma, \bar\sigma) \Leftrightarrow (\bar \sigma,\sigma)$, 
  as well as processes like 
  $(\sigma, \bar\sigma) \Leftrightarrow ({\uparrow\downarrow},0)$
  are suppressed in the large $U$ limit.
  
 The deterministic, reversible two-site update gates are then arranged in a brickwork pattern, generating the many-body dynamics in a similar way as the Suzuki-Trotter approximation of quantum evolution~\cite{Vidal-2004, Trotter-1959, Suzuki-1976}. 
 This construction allows for an efficient simulation as well as for an exact computation of certain quantities~\cite{RCA2,RCA3,Buca-2021,Klobas-2021,Penc-future}.
 The RCA model is then simulated in two ways: by sampling uniformly distributed random initial particle configurations,
 and by applying the TEBD algorithm directly to the joint classical probability distribution over the configurations (see Supplemental Material for more details~\cite{suppmat}). The particle sink is simulated by taking out particles randomly at the first site with probability $0<q\leq 1$. 

The cellular automaton model can capture some of the two-front features appearing in the probability densities.
The fast  front  emerges as a result of the reduction in the density of right-moving single fermions.
The second, slow front can be understood through the dynamics of double occupancies, as follows:
for strong interactions,  double occupancies (and   empty sites)  move via exchanging their position with 
single fermions, a first-order process in $J$.   
Due to the density imbalance between left-mover and right-mover fermions, induced by the sink, 
double occupancies start to drift away from the sink with velocity $v_\mathrm{drift} < v_0$. This drift and the 
consequent disappearance of doublons creates the second, slow front. Within the simple RCA theory, 
the drift velocity of the second front is related to the total current, $I$, flowing out from the system, and is given by
\begin{equation}
v_\mathrm{drift} = \frac{2 I }{ v_0 + 2 I} v_0.
\end{equation} 
This dependence is indeed confirmed by  comparison with our quantum simulations (see Ref.~\cite{suppmat}).

\begin{figure}
    \centering
    \includegraphics[width=1\columnwidth]{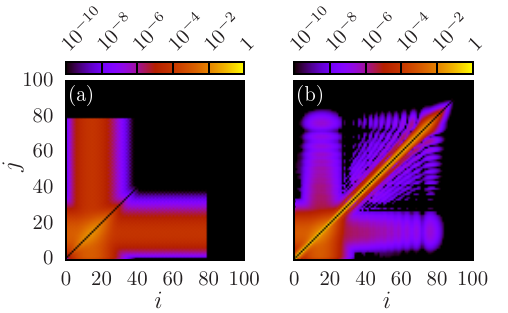}
    \caption{Mutual information \eqref{eq:MI} for the RCA model \textit{(a)}, and for the Hubbard model at interaction strength $U = \infty$ \textit{(b)}, at time $t J=80$. In the quantum case, the features similar to the cellular automaton are developed for $i\lesssim 30$ or $j\lesssim 30$. The diagonal signal for $i,j \gtrsim 30$ resembles the non-interacting case, but long-ranged interference fringes are exponentially damped.}
    \label{fig:MI_CA_inf}
\end{figure}

The  classical RCA model can also explain the mutual information features associated with  the second, slow front. 
In Fig.~\ref{fig:MI_CA_inf}.a,  we  display  the two-site mutual information computed  for the RCA model with an imperfect sink, 
$q = 0.525$. 
The long-ranged correlations following the slow front  at sites $i,j \lesssim 30$ are present in both cases.
In the RCA, a small but nonzero mutual information is also present if one of the two sites is beyond the slow front while the second is between the two fronts. This feature also appears in the quantum system in   the $U=\infty$ limit 
(see  Fig.~\ref{fig:MI_CA_inf}.b). The RCA  is, however, unable to explain the mutual information jet associated 
with the fast front.

\emph{Effective quantum theory. –}
To understand the jet feature  appearing in the large $U$ limit, we
map the $U=\infty$ Hubbard model to the product of a chain of \emph{spinless} fermions and a chain of $s={1}/{2}$ auxiliary spins~\cite{Kumar-2009,Tartaglia-2022,suppmat}. The original $\lbrace \ket{0},\, \ket{\uparrow},\, \ket{\downarrow},\, \ket{\uparrow\downarrow} \rbrace$ local basis is represented by $\lbrace \ket{0 \uparrow},\, \ket{1 \uparrow},\, \ket{1 \downarrow},\, \ket{0 \downarrow} \rbrace$ within this mapping, respectively, where the number stands for the fermion filling and the arrow displays the state of the spin. After careful definition of local operators~\cite{suppmat}, the  Hamiltonian reads 
\begin{equation}
    {H}_{U=\infty} = -\frac{J}{2}\sum_{i} \left( {f}^\dag_i {f}_{i+1} + h.c. \right) {X}^s_{i,i+1} \; ,
\end{equation}
with ${f}^\dag_i$ the creation operator of a fermion at site $i$, and  ${X}^s_{i,i+1} = \left( \frac{1}{2} + \frac{1}{2} \sigma_i^z \sigma_{i+1}^z + \sigma_i^{-} \sigma_{i+1}^{+} +  \sigma_i^{+} \sigma_{i+1}^{-} \right)$
the exchange operator of the auxiliary spins.
In this model,  the fermions $f_i$ move as free fermions on a classical spin background, while the spin pattern simply  
follows the fermions' motion, as generated by exchange processes. We identify the fast particle carrying the 
quantum information jet, as the spinless fermion, $f_i$.

 While propagation of the fermions $f$ is not obstructed by auxiliary spin exchange processes, the spin chain does store 
 information about the fermion hopping processes. The resulting correlation between the auxiliary spins and the spinless fermions results in 
 a suppression of the correlator $C_{ij} = |\langle f^\dag_i f_j \rangle|^2$  by a factor of $1/4^{|i-j|-1}$, 
 as compared to that of non-interacting spinless fermions~\cite{suppmat}. 
 
 This fast propagating, non-zero value of $C_{ij}$ results in the fast off-diagonal mutual information signal observed in our quantum simulations, and yields the relation 
\begin{equation}
    I_{ij}^{U=\infty} \eqsim \frac{I_{ij}^{U=0}/2}{4^{|i-j|-1}} 
\end{equation}
with respect to the non-interacting $U = 0$  mutual information fronts, as verified  
in Fig.~\ref{fig:op_entropy}.b This relation is precisely satisfied for large $\Gamma$ values but remains only qualitatively valid for a weak sink $\Gamma \ll J$, in which case the time $\tau \sim \Gamma^{-1}$ of creating a hole exceeds the time scale of coherent hopping processes. (see also Ref.~\cite{suppmat}).
The mutual information between distant sites is thus exponentially damped compared to the non-interacting 
model, which explains the mutual information jet observed. 

Our results on the Hubbard model thus demonstrate that dissipative processes can induce simultaneous
quantum and classical correlations in infinite temperature integrable systems. In the strongly interacting Hubbard model, 
classical and quantum correlations 
 propagate both ballistically, but with different velocities. 
The classical contribution can be interpreted in terms of a simple reversible cellular automaton model,
which gives quantitative predictions even for large and intermediate interaction strengths $U/J \gtrsim 5$. 
The classical  RCA model is, however, unable  
to reproduce quantum interference phenomena associated with a fast front, propagating with the unrenormalized 
velocity of bare particles. The observed mutual information jet can be understood in terms of an infinite $U$
effective model, where a freely propagating spinless fermion emerges and carries the mutual information while propagating on an incoherent spin background.
 It is an important open question, whether the observed phenomena  appear also in non-integrable 
or other integrable models, and if it is possible to build predictive cellular automaton simulations in such
systems. Adding a disordered external potential, or generalizing the model with longer-ranged interactions may lead to localization or invalidate the simple RCA description by suppressing the drift of double occupancies. The interplay of interactions, disorder, and dissipation needs, however, further investigation.

We close our discussions by comparing our results to former studies of dynamics of dissipative low-temperature interacting spinless fermion chains~\cite{Froml-2019,Froml-2020,Wolff-2020}. That model, equivalent to the $XXZ$ spin chain, behaves very differently at the large interaction limit, as the drift of doublons is not possible in first-order processes, therefore we do not expect the emergence of two ballistic fronts, even in the infinite temperature limit. On the other hand, one could also study the dissipative dynamics of the Hubbard model at finite temperatures. Quantum Zeno effect, i.e. the non-monotonous dependence on the dissipator strength is expected to remain in effect regardless of the temperature, similar to the spinless model~\cite{Froml-2019,Froml-2020,Wolff-2020}. At very low temperatures, $T \ll J$, one may observe Friedel-oscillations around the sink, and the velocity of the fast front is also expected to depend on filling and interaction strength~\cite{Wolff-2020}. At such low temperatures, and half-filling the density of holes and doublons is exponentially small, while the particles' spins at different sites are correlated, which properties may strongly influence the dynamics. Performing simulations at low temperatures, however, is a much harder numerical task. As the initial state is already strongly correlated, our matrix product state based method~\cite{Moca-2022} gets more challenging. As an alternative, one could use the method of quantum trajectories with stochastic sampling, but in this case, the large simulation times $tJ \approx 80$ presented in our work are probably unreachable. 
      
\emph{Acknowledgements. –}
We thank Balázs Pozsgay, Szilárd Szalay, and Frank Pollmann for valuable discussions. 
This research was supported by the Ministry of Culture and Innovation and the National Research, 
Development and Innovation Office (NKFIH) within the Quantum Information National Laboratory of Hungary (Grant No. 2022-2.1.1-NL-2022-00004), and through  NKFIH  research grants Nos. K134983, K138606 and SNN139581, 
P.P. and M.A.W have been supported by the \'{U}NKP-23-2-III-BME-327 and \'{U}NKP-22-V-BME-330  New National Excellence Programme of the Ministry for Culture and Innovation from the source of the National Research, Development and Innovation Fund.
M.A.W has also been supported by the  Bolyai Research Scholarship of the
Hungarian Academy of Sciences.
C.P.M acknowledges support by the Ministry of Research, Innovation and Digitization, CNCS/CCCDI–UEFISCDI, 
under the project for funding the excellence, contract No. 29 PFE/30.12.2021.
T.P. acknowledges Program P1-0402, and grants N1-0334, N1-0219, and N1-0233 of Slovenian Research and Innovation Agency (ARIS),
as well Advanced Grant QUEST-101096208, of the European Research Council (ERC).
\"O.L. acknowledges financial support
by the Hans Fischer Senior Fellowship programme funded by the Technical University
of Munich – Institute for Advanced Study and by
the Center for Scalable and Predictive methods
for Excitation and Correlated phenomena (SPEC),
funded as part of the Computational Chemical Sciences Program by the U.S. Department of Energy
(DOE), Office of Science, Office of Basic Energy Sciences, Division of Chemical Sciences, Geosciences, and Biosciences at Pacific Northwest National Laboratory.

%
    
\pagebreak
\widetext
\begin{center}
\textbf{\large Supplemental material to "Loss-induced quantum information jet in an infinite temperature Hubbard chain"}
\end{center}
\setcounter{equation}{0}
\setcounter{figure}{0}
\setcounter{table}{0}
\setcounter{page}{1}
\makeatletter
\renewcommand{\theequation}{S\arabic{equation}}
\renewcommand{\thefigure}{S\arabic{figure}}
\renewcommand{\bibnumfmt}[1]{[S#1]}
\renewcommand{\citenumfont}[1]{S#1}
\section{Additional details of the TEBD simulations}
\subsection{Test of ballistic front propagation in the Lindbladian dynamics}
\begin{figure*}
    \centering
    \includegraphics[width=\textwidth]{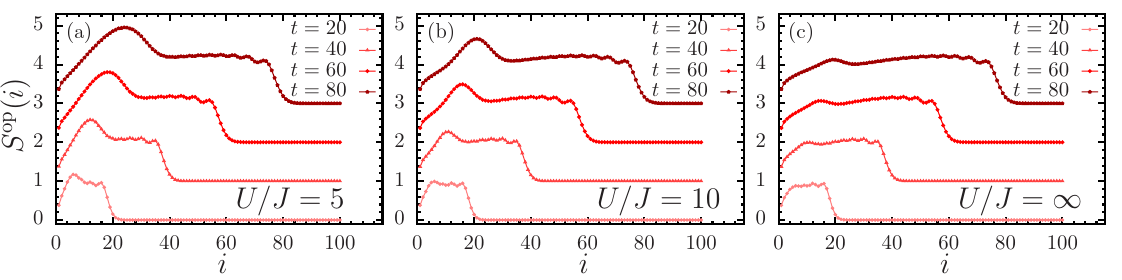}
    \caption{Operator  entanglement entropy $S^\mathrm{op}(i)$ at interaction strength (\textit{a}) $U/J=5$, (\textit{b}) $U/J=10$ and (\textit{c}) $U/J=\infty$, and dissipator strength $\Gamma/J = 0.5$ for various times. To show the double velocity of the two fronts, the profiles are shifted in the vertical direction by 1 for the different time snapshots. The velocity of the faster front is independent of the interaction strength, while the velocity of the slower front decreases with higher interaction.}
    \label{suppfig:op_entropy}
\end{figure*}

TEBD simulations of the Lindblad equation (Eq. (2) in the main text) have been performed for various interaction strengths $U/J$ including also the $U=\infty$ limit. The multiplet bond dimension of the simulations has been set to $M_{\mathrm{mult}} = 500$, and its sufficiency has been tested by checking the stability of the results against the bond dimension. Various quantities like operator entropies, occupation probabilities, and two-site mutual information profiles have been determined for snapshots at different times. Supp.~Fig.~\ref{suppfig:op_entropy} shows the operator entropy profiles $S^\mathrm{op}(i)$ for various interaction strengths and times, which profiles display ballistic propagation of the two fronts. The slower front has a finite velocity even in the $U=\infty$ limit. Supp.~Fig~\ref{suppfig:occupation} shows the probabilities of empty, singly-, and doubly occupied sites at time $tJ = 80$. The emergence of two fronts for nonzero interactions is also demonstrated by these results: behind the fast, particle-hole symmetric front the probabilities of empty and doubly occupied states are equal, while the slower front breaks this symmetry. 

\begin{figure*}
    \centering
    \includegraphics[width=0.9\textwidth]{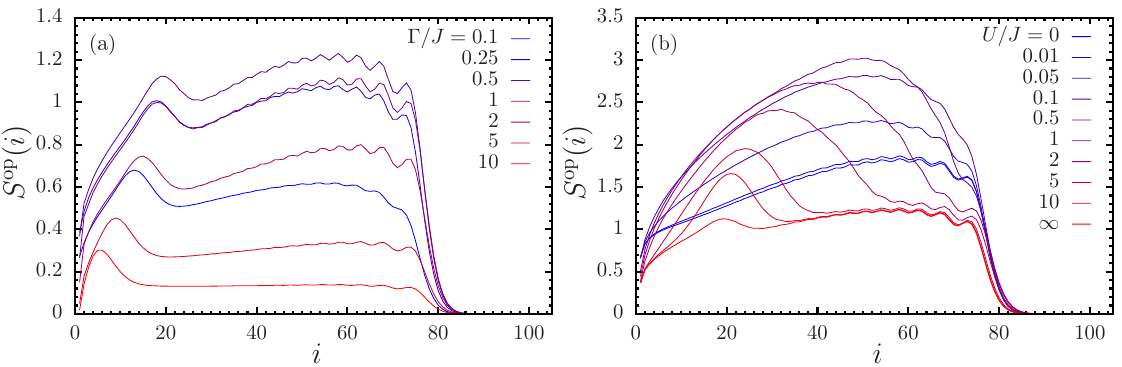}
    \caption{\textit{(left)} Operator entropy profiles at $U=\infty$ for various dissipator strengths $\Gamma$. The amplitude of $S^\mathrm{op}$ depends non-monotonically on $\Gamma$ and has a maximum at around $\Gamma \approx 0.5$. For all $\Gamma$ values, the fast front is followed by a slower bump, whose velocity depends again non-monotonically on the dissipator strength. \textit{(right)} Operator entropy profiles at $\Gamma/J = 0.5$ for various interaction strengths. The curves resemble the non-interacting result for weak interactions $U/J < 0.1$, but the magnitude of the signal is increased. The two-front structure continuously emerges for $U/J \gtrsim 0.5$ where the magnitude and velocity of the second bump decreases with increasing $U$.  }
    \label{suppfig:opentropy_GammaU}
\end{figure*}
\subsection{Two fronts in the operator entropy for various $U$ and $\Gamma$ values}

In this section, we present supplemental data on the operator entropy profiles at time $tJ = 80$ for various interaction ($U$) and dissipator ($\Gamma$) strengths. The left panel of Supp. Fig.~\ref{suppfig:opentropy_GammaU} shows the operator entropy at infinite interaction for various $\Gamma$ values. Our first observation is that the qualitative shape of these curves is very similar. A fast front of velocity $v_0/J \sim 1$ is followed by a slower bump, whose velocity is not fixed but depends non-monotonously on the dissipator strength. The velocity of the second front depends on the outflowing current (see also Supp. Fig.~\ref{suppfig:CA_drift}) and the non-monotonic behavior is a consequence of the quantum Zeno effect.

The right panel of Supp. Fig.~\ref{suppfig:opentropy_GammaU} shows the operator entropy profiles at time $tJ = 80$ with dissipator strength $\Gamma/J = 0.5$ and various interaction strengths. This figure supplements Fig.2a of the main text with curves at weak interactions. Data at very small interactions $U/J \lesssim0.1$ resembles the non-interacting result but the magnitude of $S^\mathrm{op}$ is increased. For larger $U$ values we see a continuous transition to the two-front behavior. While the slower front slows down with increasing $U$, the magnitude of $S^{\mathrm{op}}$ gets also reduced, but both the velocity and the magnitude of the slow bump remain finite even at infinitely strong interactions.

\begin{figure*}
    \centering
    \includegraphics[width=1.0\textwidth]{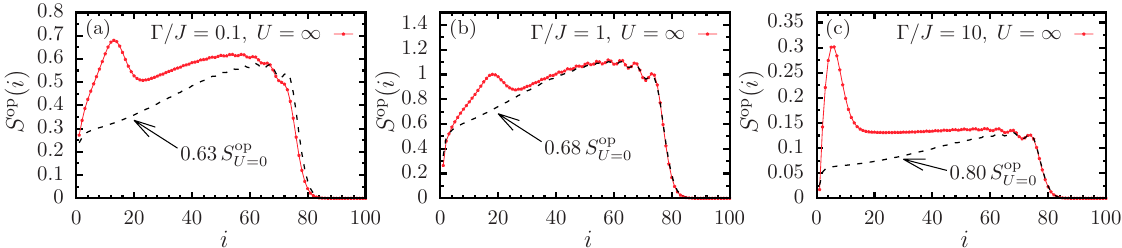}
    \caption{Operator entropy profiles at time $tJ=80$ and interaction strength $U=\infty$ for small (left), intermediate (center), and large (right) dissipator strengths. The specific values of $\Gamma$ are shown in the figures. Dashed lines indicate the operator entropy profile of the non-interacting model, multiplied by a prefactor that is determined by fitting. }
    \label{suppfig:opentropy_nonint_comp}
\end{figure*}
In Supp. Fig.~\ref{suppfig:opentropy_nonint_comp} we show selected data from the left panel of Supp. Fig.~\ref{suppfig:opentropy_GammaU} together with the non-interacting profile multiplied by a $\Gamma$-dependent prefactor that is determined by searching for the best fit. While the fast front's shape differs from the non-interacting result for the smallest $\Gamma$ value, we observe a very good agreement for intermediate and large dissipator strengths. For $\Gamma/J = 1$ the agreement is almost perfect for a large spatial range of lattice positions $i \in [40,80]$, for $\Gamma/J=10$ the agreement is perfect only in the vicinity of the front.

\subsection{Additional details of simulations of the RCA model}
The RCA model has been simulated in two alternative ways: by Monte Carlo sampling of uniformly distributed random initial configurations and by an appropriately modified variant of our TEBD code. The size of the random sample in the Monte Carlo simulations has been set to $500 000$ with a chain of length $L=100$. While the update rules of the RCA model are reversible and fully deterministic, the imperfect sink was simulated by random removal processes of probability $q$ on the first site. 

\begin{figure*}
    \centering
    \includegraphics[width=\textwidth]{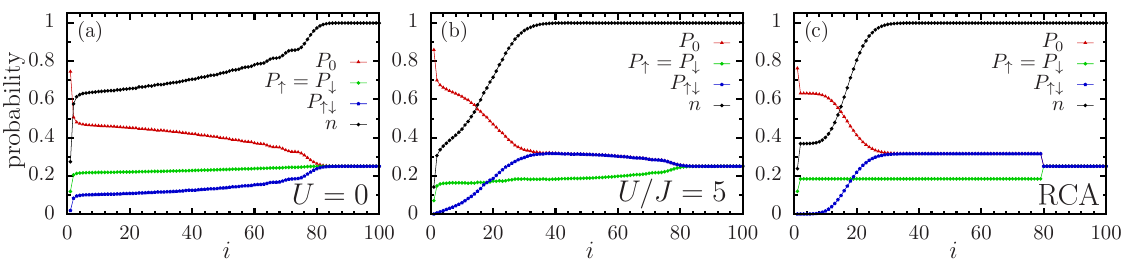}
    \caption{Position dependent  probabilities of the states $\ket{0}$, $\ket{\uparrow}$, $\ket{\downarrow}$, and $\ket{\uparrow \downarrow}$, (\textit{a}) in the non-interacting limit, (\textit{b}) at finite interaction and (\textit{c}) in the RCA model. In the non-interacting case, there is only one front, which breaks the particle-hole symmetry. At non-zero interaction and in the RCA model, the particle-hole symmetry holds between the two fronts and gets broken at the slower front.}
    \label{suppfig:occupation}
\end{figure*}

Direct TEBD simulation of the RCA model needs only a slight modification compared to the quantum case: \textit{i.)} The density operator $\rho$ is restricted to its diagonal elements, thus the local dimension is decreased from $d=16$ to $d_{\mathrm{diag}}=4$. \textit{ii.)} The two site unitary gates of the Trotter-Suzuki approximation are replaced by the two site update gates of the RCA model. The probability distribution of the RCA model has been then simulated by using bond dimension $M=2000$ (see Supp.~Fig. \ref{suppfig:occupation}).  

\section{Derivation  of the $v_\mathrm{drift} (I)$ expression in the RCA model}
\begin{figure*}
    \centering
    \includegraphics[width=0.7\textwidth]{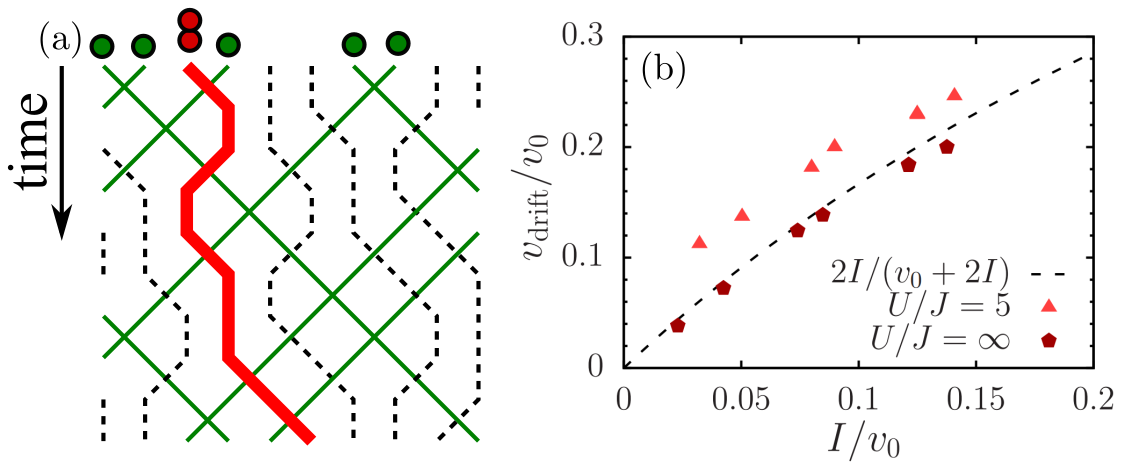}
    \caption{(\textit{a}) The motion of a double occupied site in the cellular automaton model. (\textit{b}) The velocity of the slower front as a function of the out-flowing current from the system in the RCA model, and in the Hubbard model at different interaction strengths.}
    \label{suppfig:CA_drift}
\end{figure*}

Here we present a simple derivation of the expression $v_\mathrm{drift} = 2Iv_0 / (v_0+2I)$ where $v_\mathrm{drift}$ denotes the drift velocity of double occupancies, $v_0$ is the unrenormalized velocity of fermions, while $I$ denotes the total current flowing out of the system. The key observation is that within the RCA model, the charge information of singly occupied sites travels ballistically with velocity $v_0$. The direction of this ballistic motion depends on the sublattice index of the singly occupied site: odd (even) sites travel to the right (left). A double occupancy, however, can only move if it collides with a single fermion: a collision with a left (right) moving fermion moves the double occupancy to the right (left) by one site (see Supp.~Fig.~\ref{suppfig:CA_drift}). 

As the system has been initialized in an infinite temperature state, the probability of a single occupancy is initially $1/2$ both on the even and odd sublattices. Word lines of single particles bounce back from the left end of the half-infinite chain but the sink removes particles by probability $q$. Consequently, the density of right mover single fermions goes down from $\frac{1}{4}$ to $\frac{1}{4} (1-q)$ behind a fast front moving away from the sink by velocity $v_0$. As the density of left movers remains $ \frac{1}{4}$, the total current flowing towards the sink is $I = \frac{1}{4} v_0 q$. 

The position of a selected doublon is determined by the total number of single fermions left from the doublon (together with the number of removed fermions, $It$)
\begin{equation}
    x = v_\mathrm{drift} t = N_\mathrm{fermion}^{<x}= It + \frac{1}{4} v_\mathrm{drift} t(1+(1-q))= It + \frac{v_\mathrm{drift} t}{2}(1-2I/v_0) \; ,
\end{equation}
from which expression we immediately get the result $v_\mathrm{drift} = 2Iv_0 / (v_0+2I)$. This result of the RCA model agrees well with the quantum result of the Hubbard model in the $U=\infty$ limit (see SuppFig.~\ref{suppfig:CA_drift}).

\section{Construction of the effective model in the $U = \infty$ limit}
The mapping of the $U=\infty$ Hamiltonian to the form of Eq.(5) of the main text relies on the mapping of local spin $\frac{1}{2}$ fermions to a product of a spinless fermion mode and an $S=1/2$ spin,
\begin{eqnarray}
c_{i\uparrow} &=& f_i \frac{1 + \sigma_i^z}{2} \pm f^\dag_i \frac{1 - \sigma_i^z}{2} \; , \nonumber \\
c_{i\downarrow} &=& (f_i \mp f_i^\dag) \sigma_i^{+} \; , \nonumber \\
c^\dag_{i\uparrow} &=& f^\dag_i \frac{1 + \sigma_i^z}{2} \pm f_i \frac{1 - \sigma_i^z}{2} \; , \nonumber \\
c^\dag_{i\downarrow} &=& (f^\dag_i \mp f_i) \sigma_i^{-} \; ,
\end{eqnarray}
where the alternating signs are introduced for the even and odd sublattices to get a more convenient final result. The $f^\dag_i$ and $f_i$ operators create and annihilate the spinless fermion at site $i$, respectively, while $\sigma_i^z$ and $\sigma_i^{\pm}$ stand for the usual Pauli matrices at site $i$.

The Hubbard interaction term in this basis and at half filling gets the form
\begin{equation}
    U \left(n_{i,\uparrow}-\frac{1}{2} \right) \left(n_{i,\downarrow}-\frac{1}{2} \right) = \frac{U}{4} (1-f_i^\dag f_i) \; .
\end{equation}
Consequently, the interaction energy depends only on the total number of the spinless fermions. Rewriting the full Hamiltonian in this new basis, and omitting terms that change the number of spinless fermions (which terms are therefore fully suppressed in the $U\rightarrow \infty$ limit), one gets
\begin{eqnarray}
    H &=& -J \sum_{i} \left( f^\dag_i f_{i+1} + f^\dag_{i+1} f_i \right) \left( \frac{1}{2} + \frac{1}{2} \sigma_i^z \sigma_{i+1}^z + \sigma_i^{-} \sigma_{i+1}^{+} +  \sigma_i^{+} \sigma_{i+1}^{-} \right) = \nonumber \\
    &=& -J \sum_{i} \left( f^\dag_i f_{i+1} + f^\dag_{i+1} f_i \right) X^{s}_{i,i+1} \; , \label{S.eq:Heff_infU}
\end{eqnarray}
where $X^{s}_{i,i+1}$ is the spin exchange operator that swaps the spin states of site $i$ and $i+1$.

\subsection{Mutual information jet in the $U=\infty$ limit}
The form \eqref{S.eq:Heff_infU} of the Hamiltonian allows for a formally exact solution of the time-dependent Schrödinger equation for any Slater-determinant (product) initial state,
\begin{equation}
    \left| \Psi(0) \right\rangle = \left| \Psi_F(0) \right \rangle \left| \Psi_S(0) \right \rangle =  \left| n_1^0 n_2^0 ... n_L^0 \right \rangle \left| \sigma_1^0 \sigma_2^0 ... \sigma_L^0 \right \rangle  \; ,
\end{equation}
where $|\Psi_F(0) \rangle = \left| n_1^0 n_2^0 ... n_L^0 \right \rangle$ stands for the initial configuration of spinless fermions with $n_i^0 \in {0,1}$ and $|\Psi_S(0)\rangle = \left| \sigma_1^0 \sigma_2^0 ... \sigma_L^0 \right \rangle$ is the initial spin configuration. As the Hamiltonian \eqref{S.eq:Heff_infU} describes hopping processes where also a spin-swap operation is performed, the time-dependent wavefunction will have the structure
\begin{equation}
     \left| \Psi(t) \right\rangle = \sum_{n_1,n_2,\dots,n_L} \mathcal{C}_{n_1,n_2,\dots,n_L}(t) \left| n_1 n_2 ... n_L \right \rangle  \hat{X}_{n^0_1 n^0_2 ... n^0_L}^{n_1 n_2 ... n_L}\left| \sigma_1^0 \sigma_2^0 ... \sigma_L^0 \right \rangle \; ,
\end{equation}
where the operator $\hat{X}_{n^0_1 n^0_2 ... n^0_L}^{n_1 n_2 ... n_L}$ reorders the spins according to the hopping processes between the fermion configurations $\left| n_1^0 n_2^0 ... n_L^0 \right \rangle$ and $\left| n_1 n_2 ... n_L \right \rangle$~\footnote{As the operator $\hat{X}_{n^0_1 n^0_2 ... n^0_L}^{n_1 n_2 ... n_L}$ has a clear definition only in one dimension, our solution cannot be generalized to higher dimensions.}. The amplitudes $\mathcal{C}_{n_1,n_2,\dots,n_L}(t)$ are equal to the ones of the non-interacting spinless fermion model,
\begin{eqnarray}
\left| \Psi_F(t) \right\rangle &=& e^{-i H_F t} \left| \Psi_F(0) \right\rangle = \sum_{n_1,n_2,\dots,n_L} \mathcal{C}_{n_1,n_2,\dots,n_L}(t) \left| n_1 n_2 ... n_L \right \rangle \;, \quad \textrm{with} \nonumber \\
H_F &=& -J \sum_{i} \left( f^\dag_i f_{i+1} + f^\dag_{i+1} f_i \right) \; . 
\end{eqnarray}
As the model $H_F$ is quadratic, the amplitudes $\mathcal{C}_{n_1,n_2,\dots,n_L}(t)$ can be efficiently determined.

\begin{figure*}
    \centering
    \includegraphics[width=\textwidth]{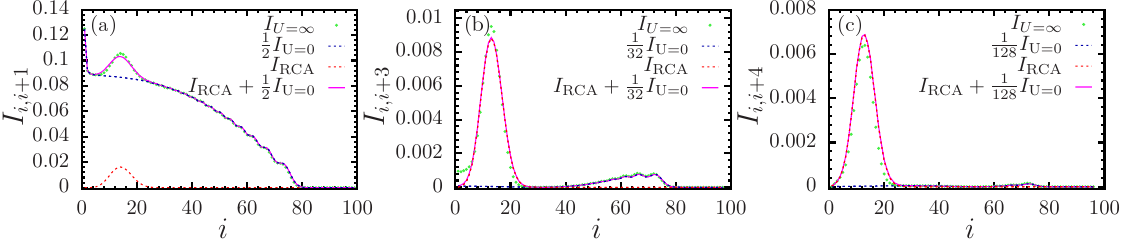}
    \caption{Mutual information profile for sites with distance (\textit{a}) one, (\textit{b}) three and (\textit{c}) four from each other. The exponentially suppressed non-interacting mutual information matches with the fast front in the mutual information. The mutual information spreading with the slow velocity is given by the sum of the suppressed non-interacting and the RCA mutual information.}
    \label{suppfig:mutinf}
\end{figure*}
Now we slightly change the initial state by making the spin configuration random. The initial total density matrix of this state is
\begin{equation}
    \rho(0) = \left| \Psi_F(0) \right\rangle \left\langle \Psi_F(0) \right| \otimes \frac{1}{2^L} \mathds{1}_{2^L} \;.
\end{equation}
The time evolution of this density matrix is given by
\begin{equation}
    \rho(t) = \sum_{\underline{n}, \tilde{\underline{n}}} \mathcal{C}_{\underline{n}}(t) \mathcal{C}_{\tilde{\underline{n}}}^*(t) \left| \underline{n} \right \rangle \left \langle \tilde{\underline{n}} \right| \otimes \frac{1}{2^L} \hat{X}_{\tilde{\underline{n}}}^{\underline{n}} \; ,
\end{equation}
where we have introduced the shortened notation $\underline{n} = {n_1,n_2,...,n_L}$ for vector of the occupation numbers. In the non-interacting spinless fermion model, whose dynamics are governed by $H_F$, the second spin-permutation part is missing from the density matrix. If we calculate now the one- and two-site reduced density matrices $\rho_i$ and $\rho_{ij}$, the diagonal elements are not affected by spin exchange, compared to the non-interacting model. The off-diagonal matrix elements are, however, suppressed by a factor of $\frac{1}{2^{|i-j|-1}}$ because tracing out the sites between $i$ and $j$ requires the spin configurations in the 'bra' and 'ket' states to be the same. As these off-diagonal matrix elements appear only in second-order corrections of the mutual information, an overall $\frac{1}{4^{|i-j|-1}}$ reduction of the mutual information is expected, if compared to the non-interacting case (see also Supp.~Fig.~\ref{suppfig:mutinf}).

\begin{figure*}
    \centering
    \includegraphics[width=\textwidth]{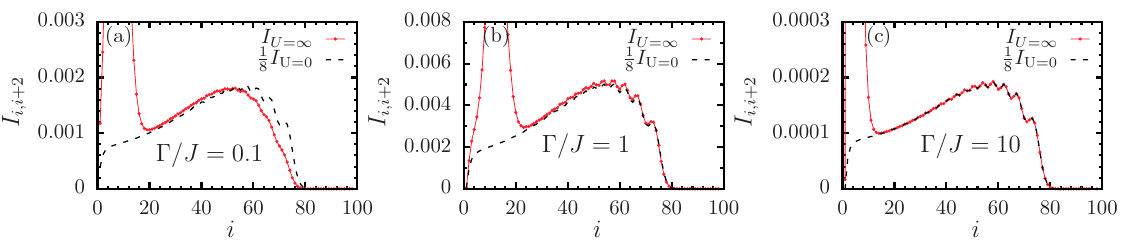}
    \caption{Next-nearest neighbor mutual information at time $tJ=80$ for infinitely strong interactions (red line and symbols) and for the non-interacting model (dashed line) for various $\Gamma$ values. In the non-interacting case, motivated by our effective theory, we multiplied $I_{i,i+2}$ by a factor of $1/8$. We observe perfect agreement for $\Gamma/J = 10$, almost perfect agreement for $\Gamma/J=1$, and stronger distortion for $\Gamma/J=0.1$.}
    \label{suppfig:mutinf_Gamma}
\end{figure*}

We note that in our theoretical construction, we suppose an initial state where charge and spin degrees of freedom are fully uncorrelated, while the charge sector is described by a (mixture) of Slater determinant(s). Then we follow only the coherent, unitary time evolution of the system. These assumptions are satisfied only, if the timescale of the dissipative creation of holes at the sink is much shorter than the timescale of coherent particle propagation. Consequently, we expect our theory to work only for large enough $\Gamma$ values. In Supp. Fig.~\ref{suppfig:mutinf_Gamma} we show the next-nearest neighbor mutual information ($I_{i,i+2}$) profiles for $\Gamma/J = \lbrace 0.1, 1, 10 \rbrace$ together with the non-interacting profile multiplied by $1/8$. The agreement at the fast front is perfect for the largest dissipator strength, still almost perfect for $\Gamma/J = 1$, but stronger distortion is observed for $\Gamma/J = 0.1$.

\end{document}